# THE FLOWING SYSTEM GASDYNAMICS
## Part 3: Saint-Venant –Wantzel's formula modern form

S.L. Arsenjev, I.B. Lozovitski[1], Y.P.Sirik

*Physical-Technical Group*
*Dobroljubova street 2, 29, Pavlograd, Dnepropetrovsk region, 51400, Ukraine*

The modern form of the Saint-Venant − Wantzel's formula for an outflow velocity of gas stream from flowing element is submitted. Taking into account of contact interaction of gas stream with the streamline surface in the form of the static head law has allowed to find the spatial−energy liaison in flowing system. The physically correct combination of mechanics of contact interaction and thermodynamics of fluid medium in one formula has allowed simultaneously to be liberated from the velocity coefficient and the discharge coefficient and polytropic process. In the new form the formula has gained the key character for computation of parameters of motion and state of gas stream in the flowing system.

## Nomenclature

- $A$    cross-sectional area of flowing element
- $\gamma_0$    weight density of fluid medium before inlet into flowing element
- $g$    acceleration of gravity
- $h$    height of current point at free fall
- $H$    general height of fall
- $k$    adiabatic exponent
- $L$    general length of flowing element
- $p_0, p_h$    quantities of pressure before inlet and on outlet of flowing element accordingly
- $p_{st}(l)$    static head in stream
- $R$    gas constant
- $T_0$    thermodynamic temperature before inlet into flowing element
- $V_{ex}$    outflow velocity (average in cross-section)
- $V_{max}$    maximum possible outflow velocity

## 1 Introduction

The development of modern groundworks of gas dynamics in the field of the flowing elements and systems is bound with necessity not only to create of the physically adequate conceptual ideas and to derive on its base of new mathematical expressions, but also revise and find the final form of some existing formulae. Saint-Venant−Wantzel's formula for determining of the outflow velocity of gas stream out of flowing element, system under specified value of pressure drop is one of the formulae, which has major value in gas dynamics of flowing systems and to which it is necessary to impart the final form.

## 2 Approach

The considered formula is a product of long-term development. Its history was started by Torricelli and Galilei experiments with the drop water stream in 1643. Further, approximately 120 years later, Borda and Du Buat have given to it the final form for mechanics and hydraulics.

After, about 60 years later (1839), Saint-Venant and Wantzel had substituted the available work of gas expansion in the conditions of mechanical and thermal isolation instead of height of free fall in the Torricelli-Galilei-Borda-DuBuat (TGBD) formula. So, SVW formula was appeared:

$$V_{ex} = \sqrt{2g\left\{\frac{k}{k-1}RT_0\left[1-\left(\frac{p_h}{p_0}\right)^{\frac{k-1}{k}}\right]\right\}} \quad (1)$$

or

$$V_{ex} = \sqrt{\frac{2}{k-1}kgRT_0\left[1-\left(\frac{p_h}{p_0}\right)^{\frac{k-1}{k}}\right]} \quad (2)$$

Writing of SVW formula in the form (1) exactly conform to substitution in TGBD formula of the available work of gas expansion in reversible adiabatic (isentropic) process. Writing of SVW formula in the form (2) imply that the outflow velocity of gas stream out of flowing element is determined by the product of a maximum possible outflow velocity of gas (outflow in vacuum) and the dimensionless radicand containing initial $p_0$

---

[1] Corresponding author.
Tel.: (38 05632) 38892, 40596
E-mail: loz@inbox.ru

and final $p_h$ pressures. In the contracted form it will match of the writing:

$$V_{ex} = V_{max}\sqrt{1-\left(\frac{p_h}{p_0}\right)^{\frac{k-1}{k}}} \qquad (3)$$

Having written TGBD formula in the form:

$$V = \sqrt{2gH\left(1-\frac{h}{H}\right)} \equiv V_{max}\sqrt{1-\frac{h}{H}} \qquad (4)$$

it is not difficult to detect that $H$ is simultaneously both difference of potentials of the propulsive energy, and the spatial longitudinal coordinate of the motion during the free fall in it.

At comparison of the TGBD and SVW formulae also it is not difficult to note, that the spatial coordinate of motion vanishes in SVW formula when the available work of gas expansion in the reversible adiabatic process is substituted instead of the height of free fall in TGBD formula. This feature is entirely typical for thermodynamics, however the usefulness of the SVW formula becomes more than doubtful for calculation of outflow velocity out of flowing element. So, both viewed formulae are non-connected with any kind of the interaction of the moving solid, fluid medium with ambient objects and mediums. And alongside with it the spatial coordinate of motion is absent in SVW formula. At first sight the unfavourable situation becomes constructive at the approach to it from the positions of contact interaction energy. The SVW formula according to such approach is the conservation equation of energy (the weight or volume density of energy more precisely) in the simplest form. The problem is to take into account the spatially - energy connection of the flowing system the physically adequately and mathematically correctly in this conservation equation. The mathematical expression for the spatially - energy connection of gas stream with wall of flowing element is necessary to have and to know, how this connection will be inserted into considered equation for the solution of the problem.

## 3 Solution

The mathematical expression of the spatially - energy connection of gas stream with wall is shown in [1, 2] in the form of static head law for gas stream in flowing element. There are simultaneously the contact interaction energy of gas stream with streamline surface and power of the counter pressure in this law. This second factor has been presented in the traditional form of the writing of SVW formula. The solution of the formulated problem is reduced to replacement of $p_h$ by $p_{st}(L)$ in the mentioned formulae (1,2). In the result we finds:

$$V_{ex} = \sqrt{\frac{2}{k-1}kgRT_0\left\{1-\left[\frac{p_{st}(L)}{p_0}\right]^{\frac{k-1}{k}}\right\}} \qquad (5)$$

or

$$V_{ex} = V_{max}\sqrt{1-\left[\frac{p_{st}(L)}{p_0}\right]^{\frac{k-1}{k}}} \qquad (6)$$

The weight-flow formula accordingly looks like:

$$G = A\gamma_0\left(\frac{p_h}{p_0}\right)^{\frac{1}{k}}V_{max}\sqrt{1-\left[\frac{p_{st}(L)}{p_0}\right]^{\frac{k-1}{k}}} \qquad (7)$$

Expressions (5,6) are the final form of the formula for the outflow velocity of gas stream out of the flowing element. The obtained expressions have generality, and SVW formula is its particular form for solution of problem concerned with the description of the gas pointwise - symmetric (spherical) expansion in gaseous medium. At the same time the mass centre of the gas is kept in rest.

## 4 Discussion of results

The obtained expression in the forms (5, 6) implies, that the actual outflow velocity of gas stream is determined as kinematic parameter of motion by its maximum possible quantity at outflow to empty space ($V_{max}$) with taking into account of energy contents of the outflowing and surrounding mediums and intensity of contact interaction of gas stream with the walls of the flowing element. Such the physical sense is much wider than sense included in it by Saint-Venant and Wantzel and consisting in the simple substitution of expression for the available work of gas expansion at reversible adiabatic (polytropic) process instead of height of free fall in TGBD formula. The new physical sense is wider as well than the traditional thermodynamic interpretation, according to which the outflow velocity is determined by difference of heat contents (enthalpies) of outflowing and surrounding mediums.

The distinguish of principle of the new formula is also what the process of the motion is presented as a mechanically irreversible process in it. This property of motion of gas stream in the flowing element is the alienable and permanent factor. The heat exchange can be absent. In this case, the out-flow process will be adiabatic. It can be absent action on the gas stream of any physical factor: heat, additional weight-flow, technical work. The factor of the contact interaction of the gas stream with the flowing element wall acts always and renders the primary and determining influence on the gas stream state and motion parameters. The expansion of the physical sense of the considered formula from narrow-thermodynamic up to the full value thermo-mechanical allows to utilize it as the effective tool for exposition both calculation of the motion and state parameters of gas stream in multiplicity of actual circumstances that is, with taking into account of the contact interaction, heat transfer, the additional weight flow, technical work, aggregate state modifications, thermo-chemical transmutations. At comparing the obtained formula with SVW formula, it is necessary to note, that they both allow to determine the gas stream velocity, firstly, on an outlet of flowing element (pipe) and, secondly, as the average weight-flow quantity that is, without taking into account of the features of its allocation on cross section of the flowing element. Along with it, it is necessary to note the difference of principle between them. Firstly, the new formula allows to take into account the influence on average velocity of the outflow of any physical factors acting on the way of the stream motion. Secondly, the new formula is free from "magic" influence to it of such conventional constant as the critical ratio of the pressure. It allows to determine the individual critical ratio for any actual flowing element with taking into account of physical factors acting on the stream. It is necessary also to note that the attempts of usage of the SVW formula for solving of technical gas dynamics problems have been giving the series of its "perfections" during last approximately 170 years.

The velocity and the weight-flow coefficients were entered by analogy with hydraulics, which define experimentally [3]. However experiment has shown, that, as against hydraulics, the quantity of the weight-flow coefficient for gas stream can be both smaller and greater unity. In one of the correctly produced experiments [4], the discharge of vessel happened much faster than on results of calculation by means of isentropic SVW formula. The polytropic process not existing in the nature for which the polytropic exponent is variable number was invented [3,5,6] by analogy with mathematical thermodynamics. Let alone that, what it is necessary to determine this exponent experimentally for each special case, this idea is situated in a complete contradiction to the molecular-kinetic theory. Introduction in gas dynamics of the isentropic process conception, entirely appropriate to SVW formula, has resulted creation of the gas dynamics tables in the form of system of the ratios devoid of the spatial coordinates of motion [7]. Not stopping on more small-sized "betterments", it is necessary to underline, that the expressions (5, 6, 7) allow correctly and precisely to present the motion of gas stream in actual circumstances and does not require changes at determining of flow parameters both in subcritical and in overcritical conditions.

## 5 Final remark

Only now, after 164 years, it is obtained the Saint-Venant-Wantzal formula form allowing for the first time rightly and exactly to determine the state and motion parameters of gas stream along the length of flowing element, system.

[1] S.L. Arsenjev, I.B. Lozovitski, Y.P. Sirik, "The flowing system gasdynamics. Part 1: On static head in the pipe flowing element,"
 http://arXiv.org/abs/physics/0301070, **2003**
[2] S.L. Arsenjev, I.B. Lozovitski, Y.P. Sirik, "The flowing system gasdynamics. Part 2: Euler's momentum conservation equation solution,"
 http://arXiv.org/abs/physics/0302020, **2003**
[3] Hutte, Handbook. Vol. 1, Sect.IV: Heat, part VI: "Gas and steam motion; subsect.a: Outflow; Outflow exponent." 15 Rus. Edition of 26 Germ. Edition, The State Scientific and Technical Publishing House, Moscow-Leningrad, p. 700-701, 1934
[4] E.M. Tseyrov "The questions of gas dynamics of air switches," Transactions of the All-Union Electrotechnical Institute. Number 60, The State Energy Publishing House, Moscow – Leningrad, p.15-17, 1956
[5] L.A. Vulis, "The gas stream thermodynamics" The State Energy Publishing House, Moscow-Leningrad, 304 p. 1950
[6] E.V. Gertz, A.I. Kudrjavtsev and all, "The pneumatic devices and systems in mechanical engineering," Handbook, The State Publishing House Mashinostrojenie, Moscow, p.11, 1981
[7] J.H. Keenan, J. Kayes, "Gas tables," John Wiley, N.Y., 1948